\documentclass[12pt,a4paper]{article}
\usepackage{amssymb,amsmath,graphicx,subcaption,cite}

\usepackage{epstopdf}
\usepackage[utf8]{inputenc}
\usepackage[english]{babel}

\begin{document}
\begin{center}{\Large {\bf Equilibrium and Nonequilibrium phase transitions in continuous symmetric classical magnets}}\end{center}

\vskip 1cm

\begin{center} {\it {Olivia Mallick and Muktish Acharyya }}\\
{Department of Physics, Presidency University, Kolkata, India}\\
\end{center}

\vskip 1cm

\tableofcontents

\newpage
\vskip 1cm

\noindent {\bf Abstract:} The magnetism is an old problem of Physics. Most interesting part of the research on magnetism
is its thermodynamic behaviour. In this review, the thermodynamic phase transitions, mainly in ferromagnetic model systems,
are discussed. The model system has a characteristic of continuous symmetry. In this context, the classical XY and Heisenberg model are chosen for discussion. With a historical survey of such phase transition observed in such models, the results of the recent (over two decades) studies are collected and reviewed. The equilibrium phase transition in such systems are discussed
to highlight the dependence of the critical temperatures on the anisotropy, dilution etc. On the other hand, 
the nonequilibrium results of such systems
driven by time dependent magnetic field, are also reviewed. We believe, this review is a modern documentation and collection of works in the field of theoretical magnetism which may be extremely useful for the researcher working in this field.

\vskip 5cm


\newpage

\vskip 2cm

\noindent {\large \bf Objectives:}\\

\begin{itemize}
    \item To discuss the previous work in the field of magnetism.
    \item To focus to the equlibrium results of thermodynamics of magnetism.
    \item The model systems are considered as having the continuous symmetry.
    \item To discuss the equilibrium phase transition of XY and Heisenberg model.
    \item To discuss the nonequilibrium phase transition of XY and Heisenberg model.
    \item To discuss the relevant technological significances.
\end{itemize}


\vskip 2 cm


\newpage

\section {Introduction:}

The magnetism in nature is an astonishing phenomenon. The material magnetism (apart from electromagnetism) has drawn the attention of the
researcher over several decades. The experimental as well as theoretical
science in the material magnetism are now a rich and well developed field
of Physics literature. 

Although, according to the Bohr-Van Leeuwen's therem, the magnetism is a quantum phenomenon, many thermodynamic aspects of material magnets can be well
described by classical models. The classical models deal with vector
(spin) which can assume any direction (generally forbidden by quantum mechanics). However, such kind of classical model of planar (
two dimensional vector) ferromagnet in two dimensions can give rise 
\cite{Kosterlitz74} to
interesting ordered phase even in the absence of long range ferromagnetic
order. Such a phase, celebrated Kosterlitz-Thouless phase, shows vortices
(pairwise vortex-antivortex) below a certain critical temperature. This
discovery of the topological phases has received Nobel prize in 2016.

The most general model of magnetic system is Heisenberg model. The thermodynamic behaviours of the ferromagnetism, antiferromagnetim, metamagnetism, canted phases, spin-glass phase all
are well explained by Heisenberg model. The anisotropy plays an important role in the transition temperature of the critical phenomena. This anisotropy breaks the special orthogonal symmetry in the planar magnetic (XY) model and SO(3) symmetry in the Heisenberg model. A huge
literature has been developed, over last several decades, regarding the phase transitions in anisotropic XY and anisotropic
Heisenberg ferromagnet. In the anisotropic planar (XY) magnet, these anisotropy in the system can be implemented in various ways.
However, in the case of Heisenberg magnet, the single site anisotropic models are investigated. The constant anisotropic 
as well as distributed anisotropic planar ferromagnetic phase (equilibrium type) transitions are studied recently and discussed
in this article.

In this review article, we mainly discussed the thermodynamc properties
of these two (XY and Heisenberg) models. Moreover, these models, driven
by the time varying  external magnetic field, give rise to the nonequilibrium phase transitions. The  
externally applied time varying magnetic field may be represented in different ways. The first kind of time dependence is the sinusoidally oscillating
(in time but uniform over the space) magnetic field. This time dependence keeps the system away from the equilibrium. 
As a result, the system undergoes a nonequilibrium kind of phase transition. The nonequilibrium phase transition in Heisenberg magnet has been disscused. Another kind of time dependence may arise when the electromagnetic wave passes through the magnetic sample. The spatio-temporal variations of magnetic field wave (the interacting part of the wave) may also give rise to nonequilibrium phase transition.  We have also discussed the interesting results of such nonequilibrium behaviours.

The whole manuscript has been organized as follows: The section-2 is devoted to the phase transition of the spin models having continuous symmetric. Section-3 contains the discussion on nonequilibrium phase transition in spin models described by the
continuous symmetry.
The technological significance is discussed in section-4. The manuscript end with some concluding remarks in section-5.

\vskip 0.5 cm

\noindent  {\large \bf }
\vskip 0.1 cm

\section {Equilibrium phase transition in Continuous Symmetric Models:}

In this section, we will discuss the phase transition in equilibrium observed in planar ferromagnetic (XY) model and Heisenberg
model. 
\label{equilibrium}

\subsection {Phase transition of the isotropic planar ferromamgnet in equilibrium:}

\subsubsection {Classical calculations:}

In this subsection will will discuss only the ferromagnetic phase transition (of classical rotor) in three dimensional XY ferromagnetic system. We will not discuss here the spin vortex related KT transition in two dimensional planar ferromagnet. The system is described by the following spin Hamiltonian:

\begin{equation}
\mathcal {H} = -J \sum_{ij} \sigma_{ix}\sigma_{jx} + \sigma_{iy}\sigma_{jy}
\end{equation}

This kind of Hamiltonian is not exactly solvable and hence recent 
Monte Carlo studies \cite{campostrini2001} are referred here to get the numerical estimates of the critical temperature and the critical exponents. 
\cite{Hasenbusch08} has estimated the critical temperature $kT_c/J=2.206$ through extensive Monte Carlo simulations.  \cite{campostrini2001}
reported the Monte Carlo estimates of the critical exponents as
$\alpha=0.0146(8)$, $\beta=0.3485(2)$, $\gamma=1.3177(5)$, $\delta
=4.780(2)$, $\nu=0.67155(27)$ and $\eta=0.0380(4)$. These set of values of the critical exponents defines the universality class of three dimensional XY ferromagnet. The domain growth and aging in planar ferromagnet has been studied 
recently \cite{Puri21}, in the presence of random field.

\vskip 0.5cm

\subsubsection {Quantum calculations:}

It may be noted here  that a quantum version of XY ferromagnet
(of spin-1/2) has
been investigated \cite{Betts68} where the Hamiltonian is 
respresented as $\mathcal { H} = -J \sum_{ij} \sigma_{ix}\sigma_{jx} + \sigma_{iy}\sigma_{jy}
=-J\sum_{ij} a_i^{\dagger}a_j+a_ia_j^{\dagger}$, where ${\vec \sigma}$ is Pauli spin matrices and $a$, $a^{\dagger}$ are
corresponding field operators coming via Jordan-Wigner transformation.
The transition temperature  
$kT_c/J \approx 4.84$ and the critical exponents $\gamma \approx 1.00$ and $\alpha \approx 0.02$ are estimated for face
centred cubic lattice. Where $\gamma$ and $\alpha$ are the critical exponents for the susceptibility and the specific heat
respectively.

\subsection{Anisotropic planar ferromagnetic model:}

The magnetic anisotropy in the planar (XY) ferromagnet plays an importane roles
in its phase. The vortex loop scaling has been studied \cite{Shenoy95} in three dimensional anisotropic XY model. The
quantum nature of the critical behaviour of
anisotropic spin-1/2 planar model ferromagnet has been studied \citenum{Su22}  with staggered Dzyaloshinskii-Moriya
kind of interaction. Geometrically frustrated XY models has been investigated 
\cite{Lach20}
to have
the emergent new kind of phase. The antiferromagnetic phases in planar (XY) ferromagnetic model on a traingular lattice with  interactions of higher order, is also reported
\cite{Lach21}, recently. Let us now report recent results on
the phase transition in three dimensional classical anisotropic XY ferromagnet. 

The anisotropic XY ferromagnet system described by the following kind of Hamiltonian:

\begin{equation}
{\mathcal H} = -J \sum_{<i,j>} (1+\gamma_{ij})\sigma_i^x\sigma_j^x + (1-\gamma_{ij})\sigma_i^y\sigma_j^y
\end{equation}
where $\sigma_{i}^x (=cos \theta_i)$ and $\sigma_{i}^y(=sin \theta_i)$ represent the components of the two dimensional vector (put in the i-th lattice position) having the magnitude unity, $|\sigma|$=1. $\theta$ represents the directional angle of orientation
of the spin vector. Classically, $\theta$ may assume any value between 0 and $2\pi$.
The anisotropy is modelled by the parameter $\gamma_{ij}$. The system restores isotropicity or conventional XY for $\gamma_{ij}=0$, and for $\gamma_{ij}=1$ the system maps to ferromagnetic XX model.
The nearest neighbour ferromagnetic interaction strength is represented by $J (>0)$. 
The anisotropy significantly governs the behaviour of magnetic systems. It would be fascinating to study the thermodynamic behaviours of the  XY ferromagnet for constant (over the lattice) anisotropy as well as for distributed
anisotropy. Here, we discuss the recent results for three different kinds of randomly (uniform, bimodal and Gaussian) distributed anisotropies \cite{mallick2023critical}.

\indent{a)\textit{Uniform distribution}}: The distribution of random anisotropy is uniform.

\begin{equation}
P_{u}(\gamma_{ij})={1\over \omega}, 
\label{uniform}
\end{equation}

\noindent here $\omega$ stands for the characteristic spread of the statistical distribution. The values of the anisotropy are randomly quenched over each lattice position. The  anisotropy may assume the values randomly within domain specified between $-\omega/2$ and $+\omega/2$. The variable $\sigma_{u}={\omega\over{2\sqrt{3}}}$ is the, variance raised to the power 1/2, of the probability density function (PDF). Hence the statistical standard deviation($\sigma_u$) is connected to the spread of the probability distribution function of the random anisotropy($\gamma_{ij}$).
 
\indent{b)\textit{Gaussian or Normal distribution}}: The random anisotropy is distributed by Gaussian formula,

\begin{equation}
P_{n}(\gamma_{ij})={1\over{\sqrt{2\pi}\omega}}e^{-{\gamma_{ij}^2\over{2{\omega^2}}}}.
\label{normal}
\end{equation}
\noindent The widely used Box Muellar algorithm has been employed to generate such Gaussian or Normal distribution. 
Here $\omega$, represents the width of the distribution. The standard deviation is denoted by $\sigma_n(=\omega)$. It
may be noted here that only for this particular kind of distribution (Gaussian) the width and standard deviation are equal.

\indent{c)\textit{ Bimodal Distribution}}: The random anisotropy follows a bimodal kind of distribution
\begin{equation}
P_{b}(\gamma_{ij})={1\over 2}[\delta(\gamma_{ij}-{\omega\over 2})+\delta(\gamma_{ij}+{\omega\over 2})].
\label{bimodal}
\end{equation}
\noindent here $\delta$ is the Dirac Delta function. Here, 
the values of the anisotropy, $-\omega/2$ or $+\omega/2$, are equally probable. The standard deviation 
($\sigma_b$) and the width of the distribution ($\omega$) are related as $\sigma_b={\omega\over 2}$.
The mean value of anisotropy $<\gamma_{ij}>$=0 for all distributions mentioned above.

\subsubsection {Role of anisotropy in the ferromagnetic phase transition:}

\begin{figure}[h]
\begin{center}

\resizebox{8cm}{!}{\includegraphics[angle=0]{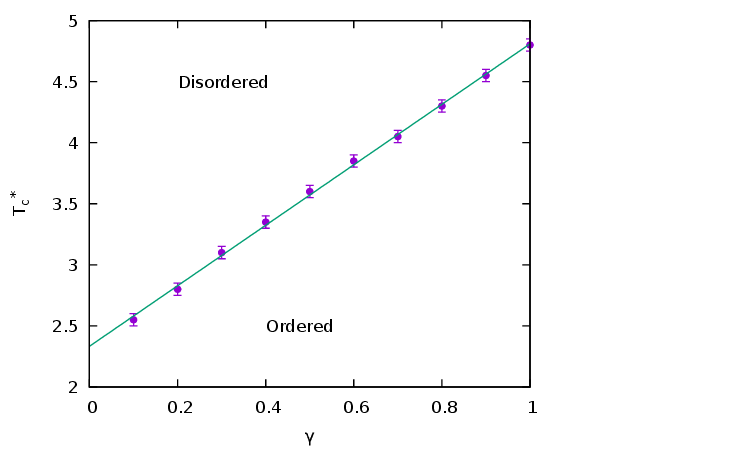}}

\caption{The finite sized transition temperature ($T_c^{*}$) is shown as a function of the magnitude of anisotropy ($\gamma$).
Here, $\gamma$ is constant over the lattice. Collected from \cite{mallick2023critical}}
\label{const-anis}
\end{center}
\end{figure}

\begin{figure}[h]
\begin{center}

\resizebox{6cm}{!}{\includegraphics[angle=0]{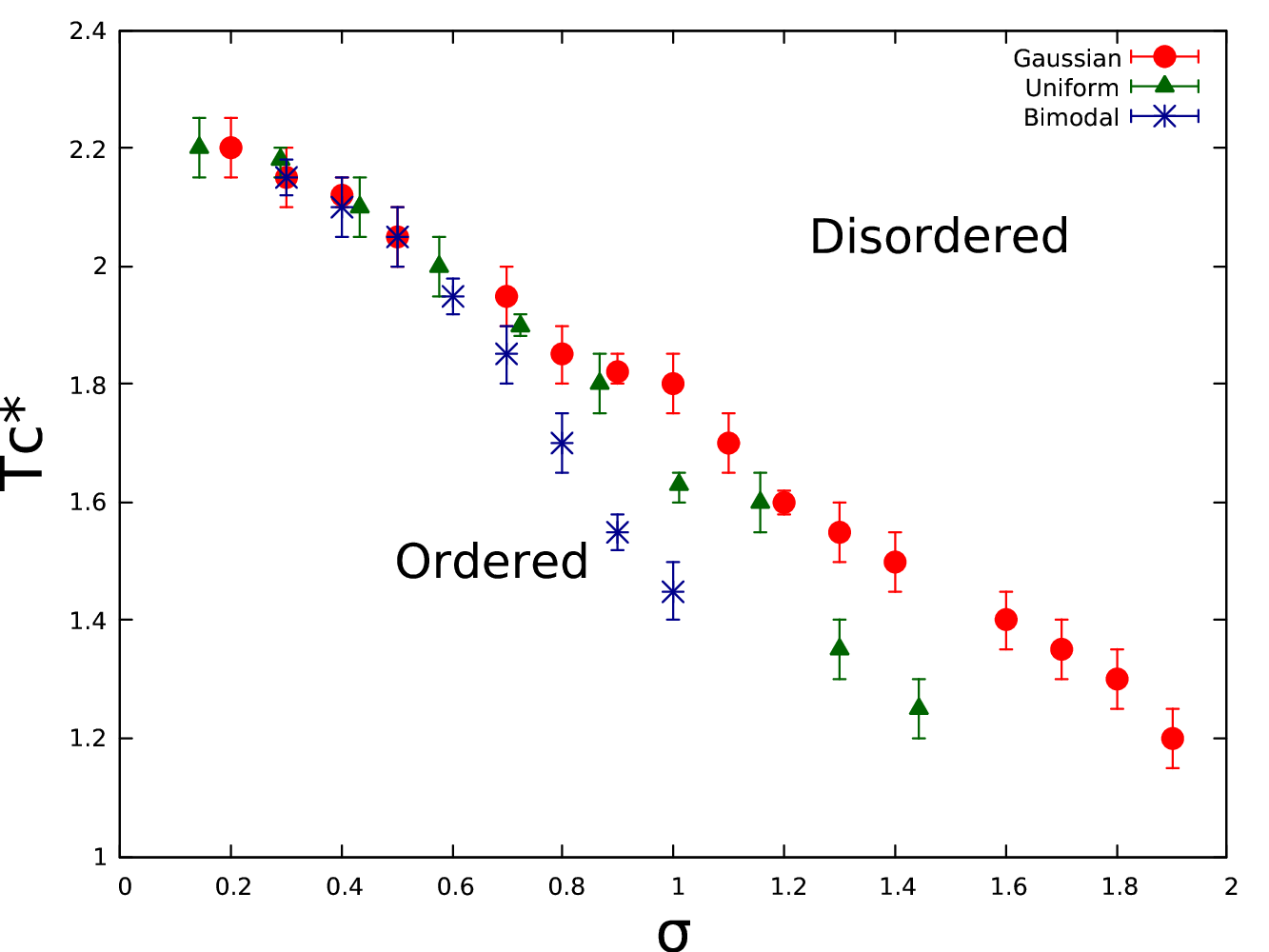}}

\caption{The finite system's transition temperature ($T_c^{*}$) is depicted as the function of the standard deviations of the various types of probability density functions of the random  anisotropy ($\gamma_{ij}$):  Uniform(Green triangle), Gaussian(Red bullet), Bimodal(Blue star). The errorbars represent the maximum error involved in determining $T_c^{*}$. Collected from \cite{mallick2023critical}}
\label{phase-diagram-dist-anis}
\end{center}
\end{figure}

The extensive Monte Carlo (MC) simulations have been performed \cite{mallick2023critical}
in three dimensional classical anisotropic XY feeromagnet using
Metropoilis algorithm. For constant anisotropy, the MC estimate of the pseudocritical temperature 
has shown a linear increase with the strength of constant anisotropy
 (Fig-\ref{const-anis}) wheras nonlinear decrase of pseudocritical temperature has been noticed 
(Fig-\ref{phase-diagram-dist-anis}) for the distributed
anisotropies. The reduction of pseudocritical temperature for distributed anisotropy may be realized just by considering
the randomly quenched anisotropy as disorder. The disorder definitely reduces the critical temperature. Whereas, the
constant anisotropy increases the critical temperature. The constant anisotropy just breaks the SO(2) symmetry
in the system which is responsible for increasing the pseudocritical temperature. All these critical behaviours are
formalized by the analysis of size dependence assuming the scaling behaviours, for the magnetisation and the susceptibility. The critical exponents (${{\beta} \over {\nu}}$ and ${{\gamma} \over {\nu}}$) are evaluated (Mallick and Acharyya, 2023).

\vskip 1cm

\newpage

What kind of critical behaviour is expected from disordered (quenched) XY ferromagnet ?
The disordered planar ferromagnet can be modelled by three dimentional XY model with randomly quenched
nonmagnetic impurity with a specified concentration.
The site diluted classical XY ferromagnet has been studied 
\cite{Santos11} by Monte Carlo simulation. The ordered ferromagnetic to disordered paramagnetic
phase transformation has been found to occur at lower temperature for nonzero impurity concentration.
The pseudocritical temperature has been studied as function of the  concentration of magnetic sites
and the results are compared with that for experimental results. The following figure shows the
results:

\begin{figure}[h]
\begin{center}
\resizebox{8cm}{!}{\includegraphics[angle=0]{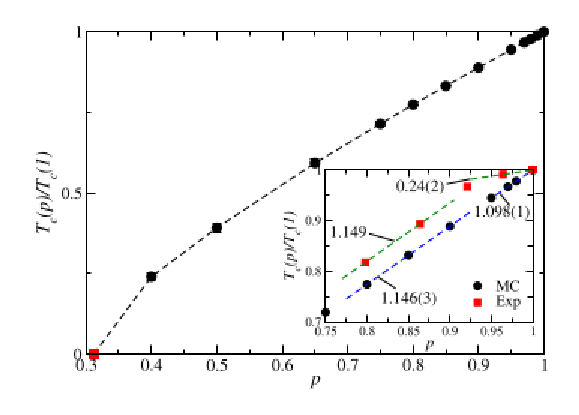}}

\caption{The effects of nonmagnetic queched impurity on the transition temperature. The transition temperature
is plotted against the concentration of magnetic sites. The percolation threshold is marked by square. The dashed line
is just joining the data points. The experimental results from \cite{Defotis08} are compared and shown in the inset is the for low concentrations.Collected from \cite{Santos11}}

\label{XY-dilute-TC}
\end{center}
\end{figure}

\newpage

\begin{figure}[h]
\begin{center}
\resizebox{8cm}{!}{\includegraphics[angle=0]{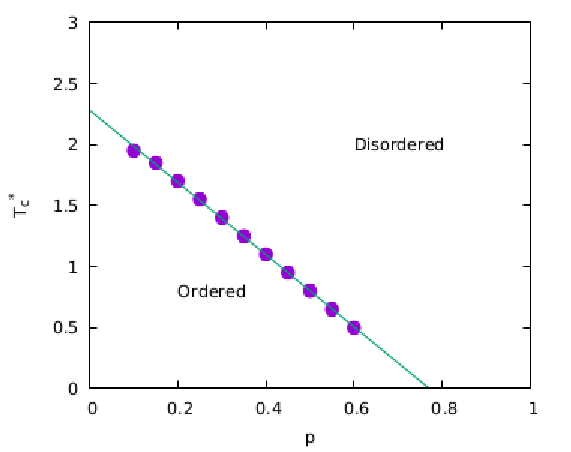}}
\caption{The pseudocritical temperature $T_c^{*}$ is plotted against the concentration $p$ of nonmagnetic impurity in a three dimensional XY ferromagnet. Collected from \cite{mallick2022critical}}
\label{XY-dilute-TC-olivia}
\end{center}
\end{figure}

However, a recent Monte Carlo study \cite{mallick2022critical} reports the linear variation 
of the pseudocritical temperature with the impurity concentration of three dimensional 
site diluted XY ferromagnet. The results are depicted in Fig-\ref{XY-dilute-TC-olivia}.

\newpage

The anisotropic XY ferromagnet has been treated quantum mechanically
to obtain the critical temperature as function of anisotropy. The $T_c$ has been calculated from magnonic dispersion relation as
$T_c={{4dJ} \over {I(0)+I(D/dJ)}}$, where $D$ is the strength of single site anisotropy and the function $I(x)$ is standard extended Watson
integral \cite{zucker2011}
The following figure (Fig-\ref{Q-phase} shows the results and the line separating the regions of ferromagnetic and paramagnetic phases. Collected from, \cite{Ma97}. The nonlinear variation in such quantum calculation is significanly different from the linear variation in the recent MC estimates \cite{mallick2022critical}. The possible reason may be the nature of
anisotropy considered in the later case.
\begin{figure}[h]
\begin{center}

\resizebox{8cm}{!}{\includegraphics[angle=0]{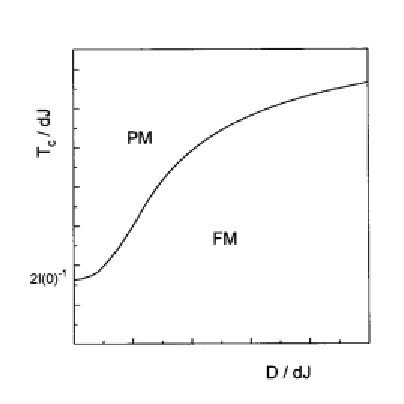}}

\caption{The dependence of the thermal critical point $T_c$  on the strength of single site anisotropy $D$.  Collected from \cite{Ma97}}
\label{Q-phase}
\end{center}
\end{figure}	

\newpage

\subsubsection {Layered XY antiferromagnet:}

So far, we have discussed the critical behaviours of anisoropic XY ferromagnet. In this section the equilibrium thermodynamic behaviour of 
the layered XY antiferromagnet will be discussed by considering 
both (ferromagnetic and antiferromagnetic) kinds of interaction in the
XY model. As a prototype, the layered XY antiferromagnet 
(or XY metamagnet) has been studied \cite{Acharyya22b}.

 The following Hamiltonian represents the XY metamagnet:

\begin{equation} H= -{J_f}\sum_{intra-planar}(\sigma_i^x \sigma_j^x+\sigma_i^y\sigma_j^y) -{J_a}\sum_{inter-planar}(\sigma_i^x \sigma_j^x+\sigma_i^y\sigma_j^y) - \sum_{i}(h_x\sigma_i^x+h_y\sigma_i^y). \label{hamiltonian}
 \end{equation}

 The  ferromagnetic nearest neighbour interaction strength within a specific plane is denoted by  $J_f ( > 0)$. To introduce the anisotropy,  the antiferromagnetic interaction strength $J_a ( < 0)$ is considered between two adjuscent planes.
  Here, the measure of relative interactions can be the ratio ($R$) of the interaction strengths $R=-{{J_a} \over {J_f}}$. The $h_x$ and $h_y$ 
  (measured in the unit of $J_f$) are the horizontal ($x$) and vertical ($y$ components of the  magnetic fields
  (applied externally) respectively. We have applied the periodic (PBC) or closed conditions of the boundaries in the all of the three spatial directions.

\begin{figure}[h]
\begin{center}

\resizebox{8cm}{!}{\includegraphics[angle=0]{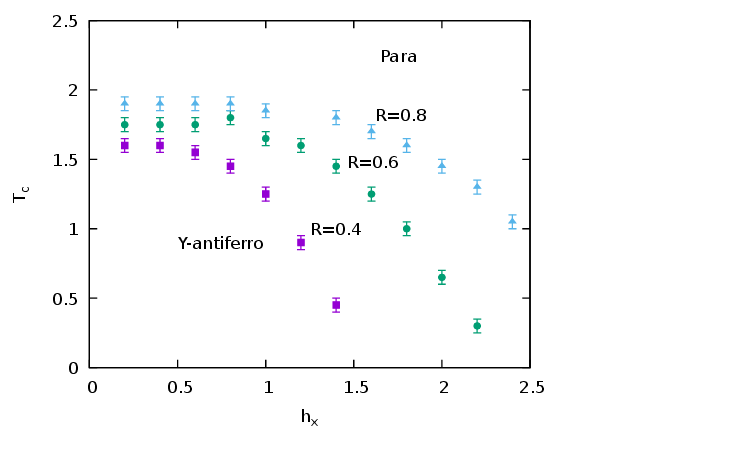}}

\caption{The phase separating lines for different values of $R$ of the layered planar antiferromagnet. Collected from \cite{Acharyya22b}}
\label{XY-meta-phase}
\end{center}
\end{figure}

The equilibrium thermodynamic behaviours of such system has been investigated by MC numerical simulation using Metropolis rule.
The system has been found to show the equilibrium phase transition.
The finite sized transition temperature has been observed to depend on the value of $h_x$ 
(we have set $h_y=0$) and $R$. The finite sized transition temperature has been predicted and evaluated by fixing the location 
of the maximum of the corresponding fluctuations of the order parameter. Fig-\ref{XY-meta-phase} shows the phase separating boundaries in the $h_x-T_c$ plane for 
different values of $R$. This thermodynamic phase transition has been formalized \cite{Acharyya22b} by finite size analysis.

\newpage

\subsection{Heisenberg Model:}
\subsubsection {Ferromagnetic phase transition in Heisenberg model: A brief historical survey }

The Heisenberg model is a good candidate to study the ferromagnetic phase transition. 
Historically, in 1969 the thermodynamic properties of two dimensional Heisenberg ferromagnet 
(spin-1/2) has been analysed \cite{Mubayi1969} by Green function technique. The analysis predicted a phase transition at critical temperature $T_c=2J/k$. The susceptibility was found to diverge at $T_c$. However, the vanishing long range ferromagnetic ordering has been observed (as predicted by \cite{Stanley1966}) 
in all (both below and above the critical temperature) which is consistent with Mermin and Wagner theorem (absence of long range ferromagnetic order 
in continuous symmetric ferromagnetic models below three dimension).

After a few years \cite{Oguchi71}, the ferromagnetic phase transition, for a general spin-$S$ two dimensional Heisenberg model has been reported. 
The variational theory predicted a phase
transformation with transition temperature ($T_r$)  given by $kT_r= (2\sigma(\sigma+1)/3)qJ$, where $\sigma$ is the magnitude of
spin, $J$ is the ferromagnetic interaction strength and $q$ is the coordination numbers. As the system cools down to  
the transition temparature $T_r$ from higher temperature, the usual static susceptibility has been found to diverge as 
$(T-T_r)^{-1}$. However, the long range ferromagnetic order is found to be zero at all finite temperatures and for all spin values. This preserves the  
the celebrated Mermin Wagner criterion. 

Later, the researchers have paid the intense attention to investigate the effects of anisotropy in the ferromagnetic behaviours of Heisenberg kind of systems. The readers may be referred a 
review article \cite{flax1970}
for a theoretical understanding of anisotropic Heisenberg model. Let us briefly mention a few of 
the important results of such studies. Historically, Landau and Binder \cite{Binder76}
reported the first Monte Carlo results of the critical properties of a two dimensional {\it anisotropic} Heisenberg ferromagnet. The anisotropy ($\Delta$) has been introduced through the
following Hamiltonian

\begin{equation}
H=-J\sum[(1-\Delta)(\sigma_i^x\sigma_j^x+\sigma_i^y\sigma_j^y)+\sigma_i^z\sigma_j^z].
\end{equation}

The existence of nonzero spontaneous magnetisation at low temperature was first reported
\cite{Frohlich77} in two dimensional anisotropic nearest neighbour Heisenberg ferromagnet.
The existence of nonzero spontaneous magnetisation at low temperature, has been concluded, as due to the presence of anisotropy.

The Monte Carlo results provided the anisotropy dependent transition temperature. The transition temperature was found to exceed the value of critical temperature  predicted for isotropic case by the method of series expansion. 

\begin{figure}[h]
\begin{center}

\resizebox{8cm}{!}{\includegraphics[angle=0]{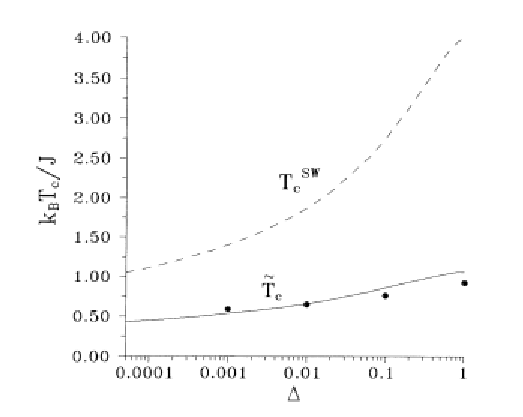}}

\caption{Critical Temperature versus anisotropy in two dimensional Heisenberg model. The spin-wave critical temperature is indicated by 
dashed line.
Collected from \cite{Serena93}}
\label{Tc-delta-heisenberg}
\end{center}
\end{figure}

The Monte Carlo simulations were performed  (with improved algorithm) (\cite{Serena93} to study the two dimensional anisotropic Heisenberg model. The anisotropy dependent critical temperature has been shown in Fig-\ref{Tc-delta-heisenberg}. The validity of spinwave analysis was found in good agreement upto the temperature $T=0.5T_c$. 

\newpage

The ultrathin magnetic film can be the realistic example of anisotropic Heisenberg ferromagnet. This
has been shown \cite{Rampini07} in anisotropic Heisenberg model with long range interaction 
investigated by extensive Monte Carlo simulations. Three distinctly different phases are identified: Namely, a ferromagnetically ordered phase, a
phase characterized by a change from out-of-plane to in-plane in the magnetization, and obviously a high-temperature paramagnetic or disordered phase.

The spin-rotation-invariant Green function theory was developed \cite{Siurakshina2000} to analyse
the behaviour of spatially anisotropic Heisenberg antiferromagnet and reported good agreement with
the experimental results in ${\rm La_2CuO_4}$ sample.

The extensive Monte Carlo simulation (with Metropolis and Wolff algorithm) has been employed to
study \cite{Freire15}
the bicritical properties of three dimensional classical Heisenberg ferromagnet with single-site anisotropy subjected to crystal field. The bicritical point, has been concluded, to belong to the three dimensional Heisenberg universality class.

\subsubsection {Metamagnetic behaviour of anisotropic Heisenberg model:Experiment versus numerical simulation}

The ${\rm FeBr_2}$ metamagnet shows many interesting phases. Experimentally observed, field induced transverse spin ordering
\cite{petracic1998}, has drawn the attention of theoretical researchers and prompted to 
visualize the effects in layered Heisenberg antiferromagnet or simply the Heisenberg metamagnet.

The continuous symmetric, Heisenberg model (with anisotropy) having competing type of spin-spin interactions
exposed to an externally applied uniform magnetic field can be represented \cite{Acharyya2000} by the energy function
\begin{equation}
  H = - J \sum_{<ij>} \vec \sigma_i \cdot \vec \sigma_j -
  J' \sum_{<ij>'} \vec \sigma_i \cdot \vec \sigma_j - 
  D\sum_i (\sigma_i^z)^2 -  \vec H \cdot \sum_i \vec \sigma_i,
\end{equation}
\noindent where the $\vec \sigma_i$ is representing a classical spin vector. This spin vector may 
assume any direction (classical picture).  For simplicity, the system has been chosen
as simple cubic type of lattice of linear size $L$. The interaction strenths
$J > 0$ and $J' < 0$ represent the intra-planar ferromagnetic and inter-planar antiferromagnetic interactions respectively. $D$ is uniaxial 
anisotropy and $\vec H$ reprsents the magnetic field applied externally.
Here also, the standard PBC were applied as the boundary conditions in all three Euclidian directions.

\begin{figure}[h]
\begin{center}
\resizebox{8cm}{!}{\includegraphics[angle=0]{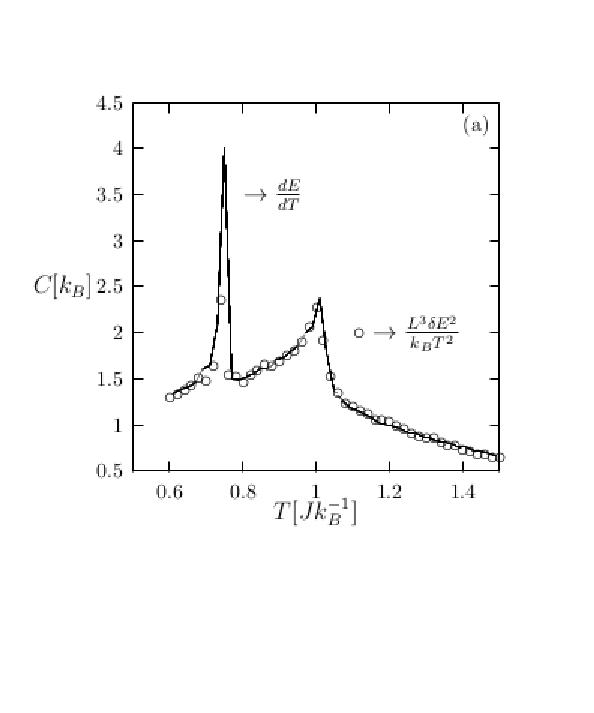}}
\caption{The specific heat of layered Heisenberg antiferromagnet. Collected from \cite{Acharyya2000}}
\label{metaphase}
\end{center}
\end{figure}

A small amount of the transversely directed externally applied magnetic field ($H_x=0.10$) has been applied additionaly with the
longitudinal field ($H_z=0.70$) to mimic the experimental set up \cite{petracic1998}. Whereas, in real experiments \cite{petracic1998},
the application of orthogonal (transverse) field is experimentally realized  by obliquely keeping the 
${\rm FeBr_2}$ metamagnetic material by a small angle measured from the
direction of applied magnetic field. The specific heat, studied experimentally, as a function of the temperature shows two distict peaks.
These two peaks indicate two distinct phase transitions. Theoretically,
\cite{Acharyya2000} the same scenario has been observed in Monte Carlo
simulations, as an evidence of the fair agreement with the experimental results. Fig-\ref{metaphase} shows the Monte Carlo results of the thermal
variation of specific heat of Heisenberg metamagnet.

Let us briefly mention the interesting findings of some related studies on Heisenberg model.
It may be mentioned here that recent quantum Monte Carlo  study \cite{Kanbur20a} the thermodynamic properties 
of a two-leg rung-disordered ladder system  shows the existence of a very special 
insensitive to the disorder strengths for the specific
heat and the susceptibility. 

The three dimensional quantum random bond Heisenberg antiferromaget was found
\cite{Kanbur20b} to be a member of three dimensional
classical O(3) symmetric universality class.

The classical Heisenberg model has been used to study \cite{Vatansever21} the magnetic phase transtion in
two dimensional Cromium based Janus MXenes. Recently, the two dimensional classical Heisenberg model has been
employed to analyse \cite{Bilican23} the strain effects on the electronic and magnetic properties of ${\rm Cr_2TaC_2}$ 
and ${\rm Cr_2TaC_2O_2}$ monolayers with reasonable agreement with the simulations results.

The recent Monte Carlo study \cite{Naskar23} reported the increase in the critical temperature, as
the intensity of the single site anisotropy of classical three dimensional anisotropic Heisenberg ferromagnet, increased.


\section {Non equilibrium phase transition in Continuous Symmetric Models:}
\label{nonequilibrium}
 \

\vskip 0.2 cm
Non-equilibrium ferromagnetic phase transitions  occur in systems far from thermal equilibrium, where the dynamics of the system play a crucial role in determining its macroscopic behavior.Unlike equilibrium phase transitions, where the system evolves towards thermal equilibrium, non-equilibrium phase transitions occur in systems driven by external forces or subjected to time-dependent perturbations. These transitions often exhibit novel and intriguing phenomena due to the interplay between dynamics, fluctuations, and symmetry-breaking. The study of non-equilibrium ferromagnetic phase transitions in continuous symmetric models provides insights into various aspects of out-of-equilibrium statistical physics. In continuous symmetric models, the order parameter characterizing the ferromagnetic phase transition is typically a magnetization vector. This vector describes the alignment of magnetic moments in the system, which can exhibit spontaneous symmetry breaking when the temperature or other control parameters cross a critical value. In this context, investigating the breaking of dynamical symmetry in continuous symmetric models during non-equilibrium ferromagnetic phase transitions is of particular interest. This phenomenon reveals how the system's dynamics affect the symmetry properties and collective behavior, offering valuable insights into the underlying physics.

\subsection{Planar ferromagnet perturbed by polarised magnetic field}
The Hamiltonian of a planner ferromagnet  can be expressed by incorporating a time-varying magnetic field that exhibits both spatial and temporal fluctuations:

\begin{equation}
\mathcal{H}(t) = -J\sum_{<ij>} \sigma_{i}^x\sigma_{j}^x + \sigma_{i}^y \sigma_{j}^y - \sum_{i} \vec{h_i}(r,t)\cdot \vec{\sigma_i} 
\end{equation}

where $J>0$ represents the strength of the exchange interaction between nearest neighboring magnetic moments, $\sigma_i^x$ and$\sigma _i^y$ denote the x and y components of the spin at site i, $\vec{h_i}(t)$ is the time-varying magnetic field at site i.The initial component of the Hamiltonian represents the interaction between adjacent spins, promoting alignment in the xy-plane and exhibiting ferromagnetic behavior. The second term accounts for the time varing magnetic field and magnetic moments at each site. The applied magnetic field exhibits both propagating and standing wave forms, characterized by their spatio-temporal variations. The propagating wave $h(r,t)= H\cos{[2\pi(ft-z/\lambda)]}$  is linearly polarized through the x-direction and  propagtes through the z-direction with its magnitude $H$.The standing wave has the form $h(r,t)=H\sin{(2\pi ft})\sin{(2\pi z/\lambda)}$ oscillating along the z-direction with the same magnitude $H$ and polarized in the x-direction. 
The framework is based  on a three dimensional cubic lattice. The lattice has periodic boundary conditions, ensuring that spins at the edges of the lattice interact with their neighboring spins on the opposite side.

MC based numerical calculation is exploited to investigate the system's behaviour \cite{Acharyya18}. The nonequilibrium order parameter 
($\vec{Q}$)components are, $Q_{\alpha} = \frac{1}{\tau}\oint M_{\alpha}(t) dt$. Where  $M_{\alpha}(t)= {1 \over {L^3}} \sum \sigma_{\alpha} (r,t)$. 
 are the instantaneous components of magnetisations and frequency $f=\frac{1}{\tau}=0.01$ taken.
The variances of $Q_{\alpha}$ (for $\alpha$-th component) are defined as 
$Var(Q_{\alpha}) = L^3(<Q_{\alpha}^2> - <Q_{\alpha}>^2)$.
 The non-equilibrium specific heat is expressed as $C= {dE \over dT}$, where $E=\frac{1}{\tau} \oint H(t)dt$ corresponds to the time averaged energy across every phase of magnetic field cycle.
\vskip 1.0 cm

\subsubsection{Partial Transition in Symmetry Breaking Dynamics}

\vskip 0.3cm

\begin{figure}[h]
\begin{center}

\resizebox{5.5cm}{!}{\includegraphics[angle=0]{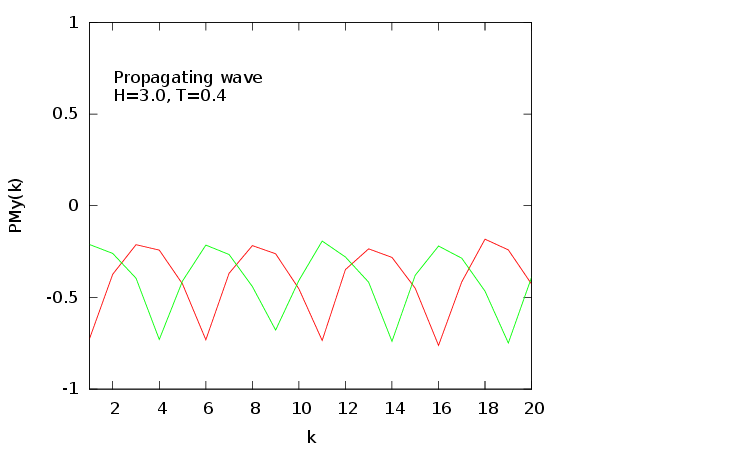}}
\resizebox{5.5cm}{!}{\includegraphics[angle=0]{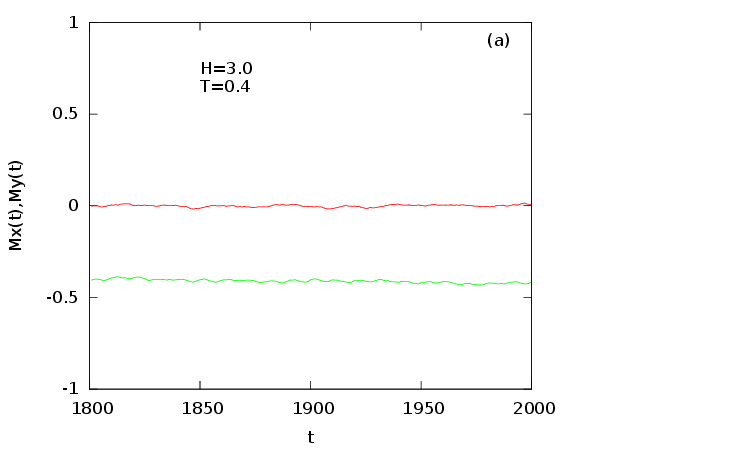}}
\resizebox{5.5cm}{!}{\includegraphics[angle=0]{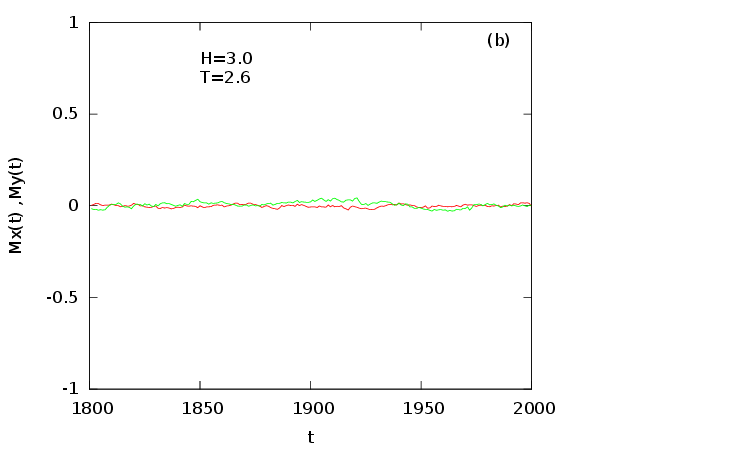}}

\caption{(a) The planar magnetization's y-component, PMy(k), is depicted against the XY plane indexed by k. distinct colours indicate distinct time instants: red corresponds ot 1900th MCSS and green corresponds to 1930th MCSS. The wave moves in the z-direction. Temporal evolution of  the horizontal (red) and vertical (green) magnetisation components  (b) at low temperature $T=0.40 J/k_B$ (symmetry broken phase) (c) at high temperature  $T=2.60 J/k_B$ (symmetric phase) at $H=3.0J$, $f=0.01(MCSS)^{-1}$ and $\lambda=10$.The diagrams illustrate the infringement of  dynamical symmetry (partial, only along y direction) in circumstance of propagating magnetic field. Collected from \cite{Acharyya18}}
\label{pw}
\end{center}
\end{figure}

A three dimensional XY planner ferromagnet exposed to a  {\it propagating or travelling} magnetic field wave with linear polarization reveals two exclusive phases that are influenced by temperature (T) and the strength of the magnetic field (H).
At low temperatures, the XY ferromagnet exhibits a phase characterized by the phase-locked motion of spin-strips aligned with the direction of the magnetic field wave propagation (z-direction). The horizontal component of the spin, on an average, turns into zero under the influnce of  linearly polarized magnetic field wave through the horizontal direction. Nevertheless, the vertical component of the spin persists nonzero on average. In contrast, at high temperatures, the spin configurations become random and structureless, resulting in both the x and y components averaging to zero.
To visualize the propagating spin mode, the y-component of the instantaneous planar magnetization is calculated for each XY plane as $PMy(k)=\frac{1}{L^2}\sum \sigma_{y}$.The  planar magnetization $PMy(k)$ is plotted against the index k, which represents different XY planes, at two different times. The plotted data, shown in Figure \ref{pw}(a), clearly demonstrates the presence of the propagating mode.

The statistical distributions of spin orientation, expressed by $\phi = \tan^{-1}\left(\frac{\sigma_y}{\sigma_x}\right)$, at a particular time , reveal the dynamically resilient spin configurations in response to a propagating magnetic field wave. At sufficiently low temperature the distribution has three modes at $\phi=0^\circ$, $\phi \approx 180^\circ $ and $\phi \approx 360^\circ$. At $\phi=90^\circ$, the distribution vanishes.
Near $\phi=270^\circ$,the distribution becomes non-zero. The distribution of angles observed confirms the existence of a predominantly y-component in the spins, while the x-component is almost negligible. In contrast, the structureless spin arrangement at elevated temperature ($T=2.60 J/k_B$) is certified by the statistically distributed spin angles having three modes of equal weightage (\cite{Acharyya18}).

The temporal variation of the magnetization components $M_{x}(t)$ and $M_{y}(t)$ are depicted  for two temperature regime  in Figure \ref{pw}(b) and (c) respectively. At sufficiently low temperature regime ($T=0.40 J/k_B$), the x-component of the instantaneous magnetization, denoted as $M_{x}(t)$, remains near zero with slight fluctuations. Conversely, the y-component, represented as $M_{y}(t)$, displays a non-zero value accompanied by fluctuations. In contrast, at higher temperatures ($T=2.60 J/k_B$), both components exhibit oscillations around zero without any appreciable deflection.

During the cooling process from high temperature, the system encounters a  breaking of dynamical symmetry partially, manifested through the behavior of the y-component of magnetization ($M_{y}(t)$). As the system undergoes cooling, the dynamic order parameter $Q_y$ encounters a transition from zero, indicating a phase of dynamic symmetry and disorder, to a non-zero value, indicating a phase of dynamic symmetry breaking and order. This transition temperature, known as the dynamic transition temperature, marks the critical point below which $Q_y$ becomes non-zero. As the maximum value ($H$) of the travelling field wave increases, the nonequilibrium transition temperature shows a decreasing pattern.
Overall, these findings highlight the dependence of the nonequilibrium transition temperature on the maximum value of the travelling field wave, as confirmed by the variance of $Q_y$ and the behavior of the nonequilibrium specific heat $C$.

It is worthmentioning that the dynamical symmetry breaking has been verified experimentally by \cite{Riego17}. The
decaying behaviour of the nonequilibrium ordered phase has also been studied experimentally in uniaxial cobalt film by \cite{Berger13}. 

\vskip 0.5cm

\noindent {\it (b) Standing wave:} 

\begin{figure}[h]
\begin{center}

\resizebox{5.5cm}{!}{\includegraphics[angle=0]{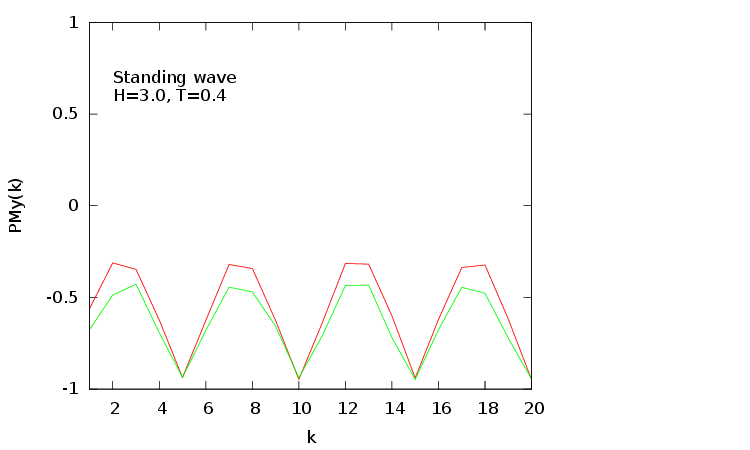}}
\resizebox{5.5cm}{!}{\includegraphics[angle=0]{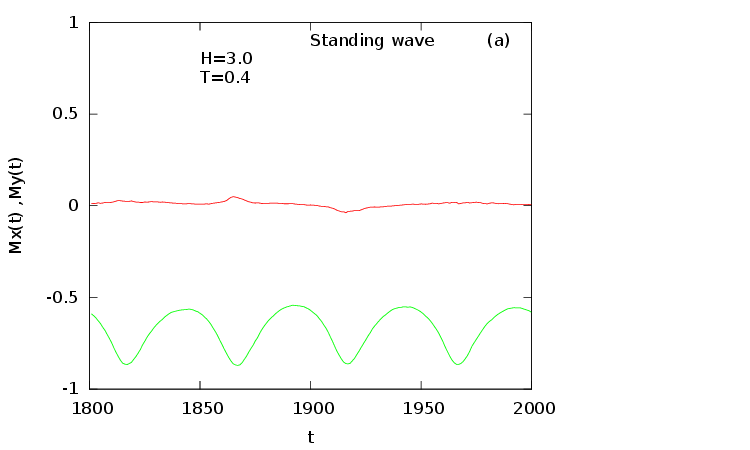}}
\resizebox{5.5cm}{!}{\includegraphics[angle=0]{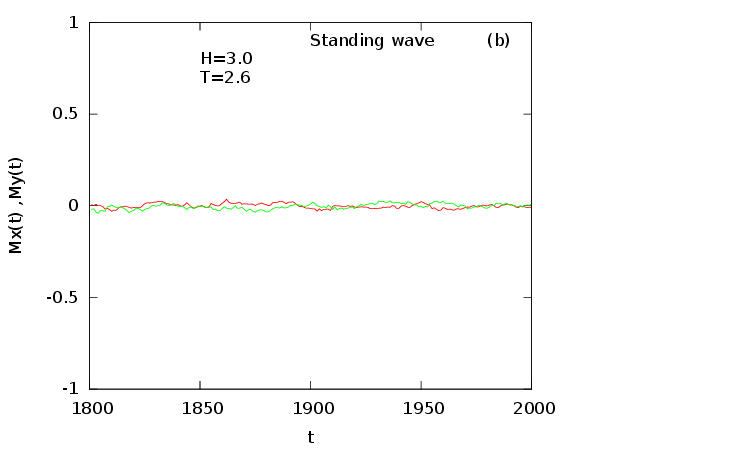}}

\caption{(a) The planar magnetization's y-component, PMy(k), is depicted against the XY plane indexed by k. distinct colours indicate distinct time instants: red corresponds ot 1900th MCSS and green corresponds to 1930th MCSS. The wave moves in the z-direction. Temporal evolution of horizontal (red ) and vertical (green) components of the magnetisation (b) at low temperature regime $T=0.40 J/K_B$  (broken-symmetric phase) (c) at high temperature  regime $T=2.60 J/K_B$ (phase with symmetry) at $H=3.00J$, $f=0.01(MCSS)^{-1}$ and $\lambda=10$.The diagrams illustrate the infringement of  dynamical symmetry (partial, only along y direction) in circumstance of standing magnetic field. Collected from \cite{Acharyya18}}
\label{sw}
\end{center}
\end{figure}

The XY ferromagnet, when subjected to a stationary magnetic field wave, displays two distinct dynamic phases that are influenced by both temperature ($T$) and the maximum value ($H$) of the field. The spin configurations are analysed in the plane formed by X and Z axes at Y=10 and H=3.0, under conditions of cold thermal condition ($T=0.40 J/k_B$) and hot thermal conditions ($T=2.60 J/k_B$). At low temperatures, a standing spin wave mode is observed, characterized by stationary spin bands.An alternative visualization of the stationary mode in an induced spin wave within the XY ferromagnet involves plotting the y-component of the planar magnetization, denoted as $PMy(k)$, as a function of the index k for the k-th XY plane. Figure \ref{sw}(a) displays these plots for two distinct time instances. 
In the case of standing magnetic field wave, the $k$ dependence of $PMy(k)$
remain unchenged in their corresponding position over time,in contrast to the behavior observed in the propagating mode.

The probability distribution function of the angles (spin vector) shows four peaks near $\phi = 0^\circ$, slightly above $\phi = 0^\circ$ , slightly below $\phi = 180^\circ$, and around $\phi = 270^\circ$. This tetramodal distribution leads to an overall negative  y-component of magnetization, with an average x-component of magnetization of zero. This represents a dynamically ordered phase with partial symmetry breaking in the y-component. At high temperatures, the spin configuration appears random and structureless. The statistical distribution of spin vector angles exhibits a symmetric and trimodal shape with peaks near $\phi = 0^\circ$,$\phi = 180^\circ$ and $\phi = 360^\circ$ of approximately equal height. This results in a vanishing net y-component of magnetization and an average x-component of magnetization of zero, representing a dynamically disordered phase.

The presence of a partially symmetry broken ordered phase at low temperatures and a symmetric phase at high temperatures in the XY ferromagnet driven by a standing wave is evident from the observations. The time-varying dynamics of the magnetization components clearly reveal the violation of dynamical symmetry. By examining the instantaneous horizontal and vertical components of the magnetisation  as time-varying function, shown in Figure \ref{sw}(b) and (c), distinct characteristics of the higher temperature ($T=2.60 J/k_B$) and lower temperature ($T=0.40 J/k_B$) phases become apparent. In the high temperature phase, a dynamically symmetric state is observed, with both horizontal and vertical components of the magnetisation varying around the zero value with almost equal importance. Conversely, in the cold thermal state, a partial breaking of symmetry is observed, specifically in the y-component $My(t)$, which exhibits an asymmetric variation around the zero line. These findings suggest the occurrence of a symmetry broken nonequilibrium transition in the perturbed planar ferromagnet under the influence of a standing/stationary  magnetic wave. Importantly, in both of the nonequilibrium phases, the mean horizontal component of the magnetization over a complete cycle of the standing wave, is precisely zero.

To explore the nonequilibrium  transition related to the symmetry destructing in a perturbed planar ferromagnet, the thermal variations of $Q_y$, $Var(Q_y)$, and $C$ were evaluated. The results revealed interesting behaviors. As the system undergoes a cooling process starting from elevated temperatures, $Q_y$ exhibited a nonzero value near a transition temperature. Notably, this transition temperature decreased with increasing values of the applied field amplitude $H$. Furthermore, sharp peaks were observed in $Var(Q_y)$ and $C$ in the vicinity of critical temperatures, serving as indicators of the phase transitions. Overall, these findings suggest that For increasing field amplitudes ($H$), the transitions come about  at lower temperatures.

Due to limited computational resources and time, a crude estimation of the transition temperatures is obtained from the locations of the peaks in $Var(Q_y)$ for different system sizes (L=20, 30, and 40) with a fixed wavelength for both the cases {\it(a)Propagating wave} and {\it(b)Standing wave}(\cite{Acharyya18}). Remarkably, The peak positions remain consistent across different values of L, while the height of the peaks increases with higher L, indicating the development of critical correlations.

\vskip 1cm

\subsubsection {Comprehensive Phase Boundary}

\begin{figure}[h]
\begin{center}

\resizebox{8.5cm}{!}{\includegraphics[angle=0]{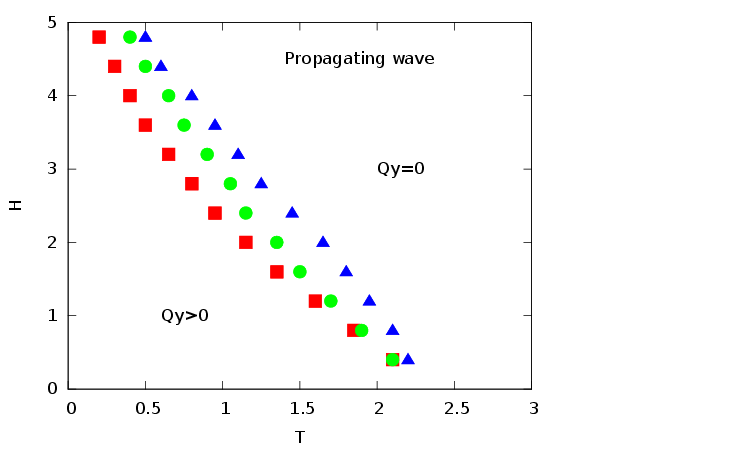}}
\resizebox{8.5cm}{!}{\includegraphics[angle=0]{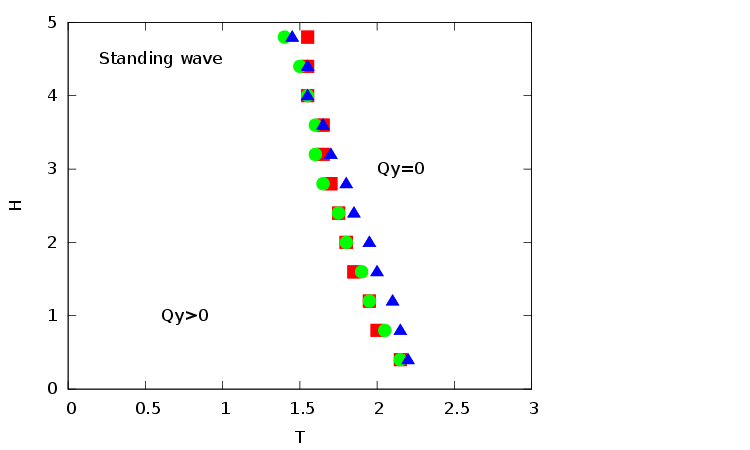}}

\caption{The sketch of phase separating lines for (a) Travelling wave and (b) Stationary wave.  Different values of wavelengths $\lambda $ are denoted by different symbols,  $\lambda $= 20(
square), $\lambda $ = 10(bullet)  and  $\lambda $ = 5(triangle). Collected from \cite{Acharyya18}}
\label{drivenXYphase}
\end{center}
\end{figure}

The dynamic transition temperatures in the system were influenced by the wavelength of the propagating magnetic field wave, while keeping the field amplitude fixed. Longer wavelengths of the propagating wave led to a transition from order to disorder at lower temperatures. A phase diagram was framed to represent the dependence of the non-equilibrium transition temperature on the maximum value of the magnetic field and wavelength of magnetic field. The phase boundaries were determined for different wavelengths, and it was observed that for longer wavelengths, the phase boundary contracted towards lower temperatures and field strengths Figure\ref{drivenXYphase}(a). In contrast, the phase boundaries in the case of a standing magnetic wave Figure\ref{drivenXYphase}(b) did not show a clear variation with wavelength. Further investigations are needed to better understand the behavior of the phase boundary in the case of a standing wave.

A qualitative explanation for this observation can be offered by taking into account the wavelength as a gauge of the exposed region where spin flips occur. With a larger exposed area, lower temperature and weaker field intensity would be sufficient to generate dynamic order within the sample. However, this explanation is inadequate to fully understand the convoluted dependence of the phase separating line on wavelength in the case of a standing/stationary wave. Thus, additional research is needed to gain a deeper insight into this behavior of the phase boundary.


In the scenario of a stationary magnetic field wave, the dynamic phase at lower temperatures manifests as a stationary spin wave configuration.  Such a nonequilibrium  kind of transition with broken symmetry  was observed, identical to the case of a travelling magnetic field wave,. The dynamic phase separating line was delineated, which displayed no significant reliance on the wavelength($\lambda $) of the stationary  field wave. Nevertheless, akin to the propagating wave case, the nonequilibrium transition temperature was found to converge toward the equilibrium  transition temperature (approximately 2.20 $J/k$) as the field amplitude approached zero, as reported by \cite{campostrini2001}.
The nonequilibrium phase transition in the  three-dimensional XY ferromagnet emerges due to the intricate interplay between the intrinsic  time scale of the propagating magnetic field wave and the inherent relaxation time of the sample. As the ferromagnetic sample cools down from a hot thermal state, it progressively loses its ability to keep pace with the driving field, primarily due to an increasing intrinsic time scale. This leads to a lagging of phases between the system's response and the time-varying perturbation, ultimately resulting in dynamic ordering. Since the field polarization aligns with the horizontal direction, the spin vector become unable to fully align with the field, causing the vertical component to possess a nonzero mean value while the horizontal component diminishes. Transitions occur at lower temperatures as the field intensity increases. The nonequilibrium effects arise from the contest between the distinct characteristics of the driving field and the inherent relaxation dynamics of the sample.

This research study extends an invitation to experimental researchers to explore the impact of  dynamics of the spins in ferromagnetic polycrystals, such as compounds of Iron and Zinc. These materials can be effectively modeled as classical XY systems with diluted sites and superexchange interactions. Investigating the reaction and thermodynamic behavior of ferromagnetic samples under intensified optical environment holds significant relevance in the realm of spintronics. By conducting experiments on these systems, valuable insights can be gained into the dynamics and potential applications of ferromagnetic materials in spintronics research.

\subsection{Anisotropic Heisenberg Ferromagnet in External Magnetic Field}

When classical  Heisenberg ferromagnetic system with anisotropy is perturbed by an external magnetic field,the  Hamiltotian can be expressed as,
\begin{equation}
\mathcal{H } = -J \sum_{<ij>}\vec \sigma_i \cdot \vec \sigma_j -D \sum_i (\sigma_i^z)^2 - \vec h \cdot \sum_i \vec \sigma_i 
\end{equation}
\noindent where $\vec \sigma_i $  represents a three dimensinal classical spin vector of magnitude 
unity ($\sigma_{ix}^2+\sigma_{iy}^2+\sigma_{iz}^2=1)$ quenched at the i-th lattice point of the sample. The three dimensional classical vector of the rotor,
$\vec \sigma_i$ can assume any  direction in the vector space. The exchange interaction term (the first term) describes the tendency of neighboring spins ($<ij>$) to align with each other, with J representing the strength of this interaction. The second term accounts for the anisotropy energy, favoring spin alignment along the z-axis, with D indicating the strength of this anisotropy.
It is pertinent to note here that when $D = 0$, the framework approaches to the isotropic Heisenberg limit, however as $D$ grows to infinity, the system approaches the Ising limit.
The third term (Zeeman term) depicts the system's interaction with an imposed time-varying magnetic field ($\vec h$). The field components vary sinusoidally in time, as defined by $h_\alpha = h_{0\alpha} \cos(\omega t)$, where $h_{0\alpha}$ denotes the amplitude having frequency $f$($f=\omega/2\pi$).

Monte Carlo simulations were employed to study the described model. The equilibrium spin configuration at temperature T was obtained by gradually cooling the system from a random initial configuration. The simulations utilized the Metropolis algorithm with a random updating scheme. Further details on the simulation method can be found in standard textbooks. The simulations were performed on a 3D isometric sample with a linear system size of L=20, PBC were imposed in all of the three spatial directions.
The $\alpha$-th component of  magnetisation at a time instant
 (per lattice site) 
is formulated as $m_\alpha(t) = \sum_i {{\sigma^i_\alpha} \over {L^3}}$.

After integrating the simultaneous magnetization components across an entire cycle of the oscillation magnetic field, the steady value of the dynamical order-parameter components have been achieved: 
$Q_{\alpha} = {1 \over {\tau}} \oint m_{\alpha}(t) dt$.  The vector addition of the three components provides the overall dynamical order parameter Q: 
$\vec Q= iQ_x + jQ_y + kQ_z$. Additionally, the instantaneous energy is given by  the Hamiltonian (Equation-10).
The time averaged  energy is defined as the integrated Hamiltonian over an
entire cycle of oscillation. The thermal gradient of energy (the specific heat) is defined as  $C  = {{dE} \over {dT}}$ (calculated by three point central difference formula). 

\vskip 1 cm

\subsubsection {Longitudinal and transverse dynamic transition}


Under the influence of a magnetic field that oscillates sinusoidally (with amplitude $h_{z}^0$ parallel to the anisotropy direction) an axial transition occurs in the dynamical order parameter component $Q_z$. At high temperature, the magnetic hysteresis loop ($m_z-h_z$) is symmetric, resulting in $Q_z$ being zero. However, at low temperature, the loop becomes asymmetric, leading to a nonzero value of $Q_z$. Similar behavior is observed when a magnetic field is put orthogonal to the direction parallel to the anisotropy ($x$-direction), where the $m_z-h_x$ loop becomes asymmetric at low temperature while the $m_x-h_x$ loop remains symmetric. These transitions indicate a dynamic symmetry breaking from a symmetric to a symmetry broken phase as the temperature decreases. The off-axial transition is characterized by a 'marginally symmetric' behavior in the $m_z-h_x$ loop at higher temperatures, where the loop lies close to the $h_x$ axis. This transition from 'marginally symmetric' to asymmetric occurs as the temperature decreases. Unlike longitudinal transitions, the transverse transitions unable to revert the z-component of spin through the application of a field perpendicular to the anisotropy direction, resulting in a marginally symmetric loop. These observations highlight the importance of anisotropy for the occurrence of dynamic symmetry breaking within the framework of the conventional Heisenberg model in classical physics. \cite{Acharyya03}.

\begin{figure}[h]
\begin{center}

\resizebox{8cm}{!}{\includegraphics[angle=0]{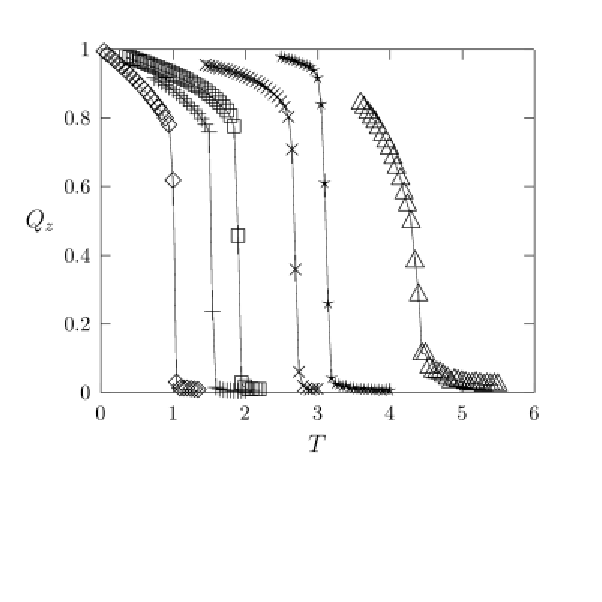}}(a) 
\resizebox{8cm}{!}{\includegraphics[angle=0]{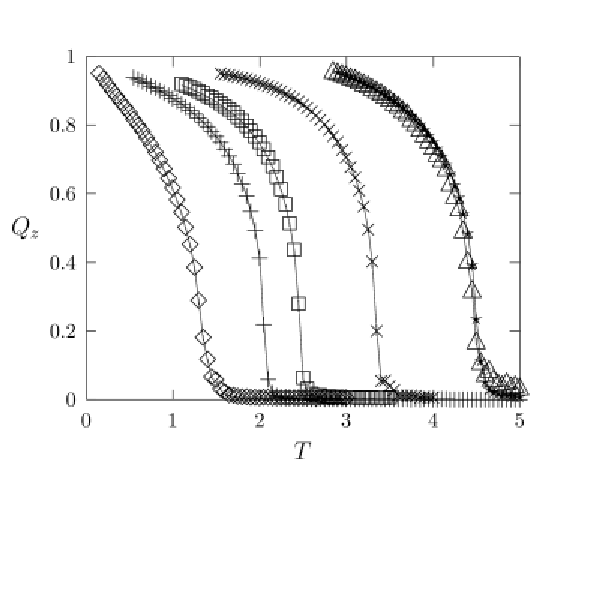}}(b)
\caption{Thermal ($T$) dependence of the z component of the order parameter  ($Q_z$) for various intensities of single site anisotropy (D): D=0.50(diamond), D=2.50(plus), D=5.00(square),D=15.00(cross) and D=400(triangle). Figure(a) corresponds to axial ($h_{z}^0=0.50J$) dynamic transition and figure (b) corresponds to off-axial ($h_{x}^0=0.50J$) dynamic transition. For both the cases the thermal dependence of nonequilibrium ordering parameter in  conventional Ising system (with $h_{z}^0=0.50J$ and $f=0.001(MCSS)^{-1}$) are depicted by (star). Collected from \cite{Acharyya03}.}
\label{axial-offaxial}
\end{center}
\end{figure}

To explore how the critical temperature relates the intensity of anisotropy ($D$) in both longitudinal type and transverse kind of transitions, the thermal dependence of the z component of dynamic order parameter ($Q_z$) was analyzed across various $D$ values. In the axial transition case \ref{axial-offaxial}(a), it was found that at small values of $D$ (for example, 0.5 and 2.5), the transition undergoes a discontinuous change, whereas at large values of $D$ (such as 5.0 and 15), the transition occurs continuously. For the Ising limit($D \rightarrow \infty$), the longitudinal transition was also observed and compared to the Ising model, showing a continuous transition occurring at a similar temperature. Similarly, in the off-axial transition \ref{axial-offaxial}(b), it was observed that the transition temperature moves towards higher values, with increasing $D$, and the nature of the transition is of continuous type for all anisotropy values. The comparison of the transition at $D=400$ with the Ising model demonstrated their similarity in terms of a continuous transition occurring at approximately the same temperature. The effect of intensty of the  anisotropy on the transition temperature, is attributed how the anisotropy aligns the spin vector along its direction, requiring higher thermal fluctuations to break the symmetry. There is a distinction in the behavior of the transition between longitudinal and transverse scenarios, where the longitudinal transition demonstrates a discontinuous nature at comaratively smaller values of the anisotropy  and develops into continuous for higher $D$ values, while the transverse transition remains continuous throughout.

\vspace{0.75cm}

 As the anisotropy approaches infinity, the characteristics or dynamics of the system differs depending on whether it is an axial or off-axial transition. When considering the scenario of longitudinal transition, the transition temperature in the Heisenberg ferromagnetic model with infinite anisotropy varies compared to that of the Ising model. Despite both models exhibiting the same transition temperature during equilibrium transitions, there is a disparity in the nonequilibrium transition temperatures. This is because the system remains in a nonequilibrium state due to the sinusoidally oscillating magnetic field applied in the z-direction, preventing it from becoming an conventional Ising system even when approaching the limit of enormously large anisotropy.
In contrast, when it comes to transverse transitions, the critical temperatures in both the infinitely anisotropic Heisenberg ferromagnet and the Ising ferromagnet coincide perfectly. This is because the applied magnetic field, that oscillates sinusoidally in the x-direction, has no impact  when anisotropy approaches to infinity. In this scenario, the infinitely anisotropic Heisenberg ferromagnet effectively maps onto the Ising ferromagnet in a state of  statistical and thermal equilibrium when the field is perpendicular to the direction of easy axis. Consequently, the nonequilibrium transition in the  Heisenberg model   with infinite ansiotropy  perturbed by an transverse field and the equilibrium Ising ferromagnet in absence of applied magnetic field yield the identical results.

It is noteworthy to highlight a significant aspect concerning the dynamics employed in this simulation.Due to the non-commutation between the spin component and the Heisenberg Hamiltonian, the spin component displays inherent dynamics. This aspect was considered in a study that investigated the transport properties and the structure factor in the planar ferromagnet, highlighting the influence of the intrinsic dynamics on the system's behavior.

\subsubsection{Elliptically polarized magnetic field: Multiple dynamic transition}


If we consider an externally applied magnetic field that possesses elliptical polarization and is oriented within the X-Z plane, we can express this mathematically as follows:,
\begin{equation}
\vec h = \hat x h_x + \hat y h_y + \hat z h_z\\
= \hat x h_{0x}{\rm cos}{(\omega t)}+ \hat z h_{0z}{\rm sin}{(\omega t)}.
\end{equation}
where,
\begin{equation}
{{h_x^2} \over {h_{0x}^2}} + {{h_z^2} \over {h_{0z}^2}} = 1,
\end{equation}
Typically, $h_{0x}$ and $h_{0z}$ have different values. However, if we assume $h_{0x} = h_{0z} = h_0$, the magnetic field will exihibit circular polarization, satisfying the condition $h_x^2 + h_z^2 = h_0^2$. 
The externally applied magnetic field and the intensity of anisotropy ($D$) are quantified in the units of $J$. The simulation method is implemented on an isometric 3D system with a linear size of $L=20$ and employed  PBC in each of 3 spatial dimensions.

 The selection of the field amplitudes and frequency involved a meticulous search process to ensure their appropriateness for the study.The components of magnetic field amplitude  were calibrated at  $h_{0x} = 0.30$ and $h_{0z} = 1.00$. A  frequency of $f=0.02$ was chosen, leading to 50 MCSS encompassing one complete cycle which corresponds to the temporal period ($\tau$) of the time varying magnetic field. The  uniaxial anisotropy strength was chosen as D = 0.2, which remained constant throughout the investigation. It is worth noting that this specific value of D was prudently elected to ensure these intriguing results. However, diverse values of D are anticipated to lead to different transitions points and as D increases, the occurence of multiple transitions is expected to diminish.

\begin{figure}[h]
\begin{center}

\resizebox{4cm}{!}{\includegraphics[angle=0]{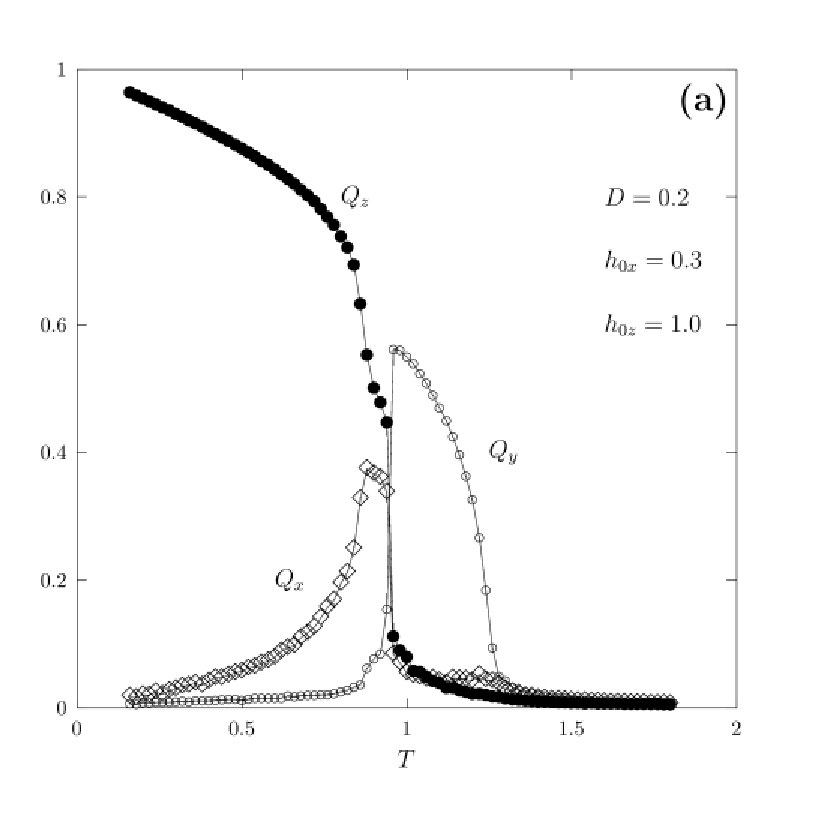}}\\
\resizebox{4cm}{!}{\includegraphics[angle=0]{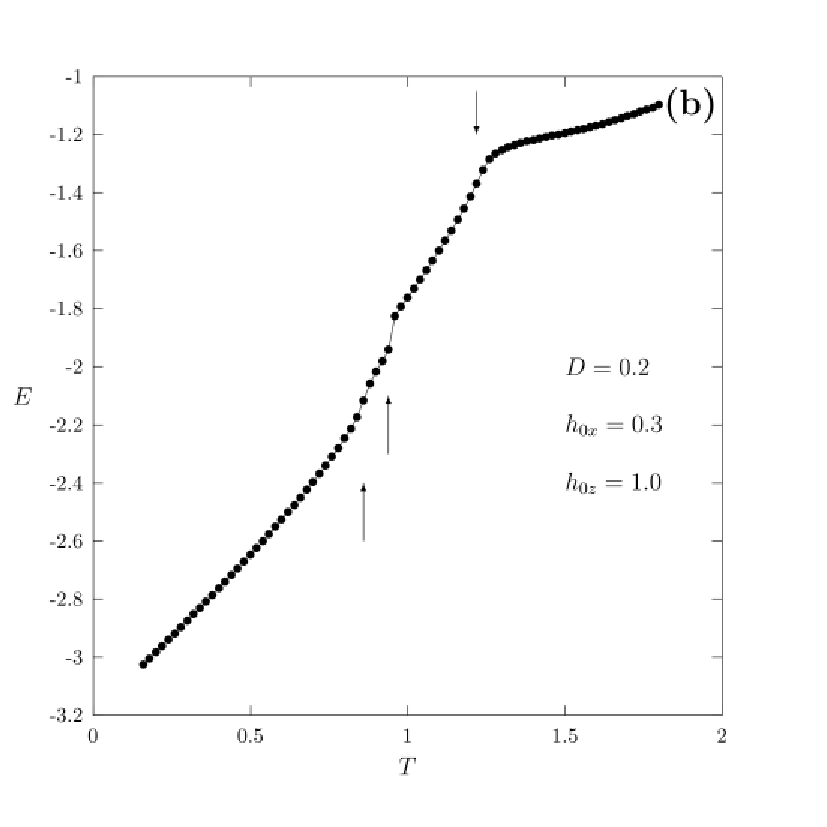}}\\
\resizebox{4cm}{!}{\includegraphics[angle=0]{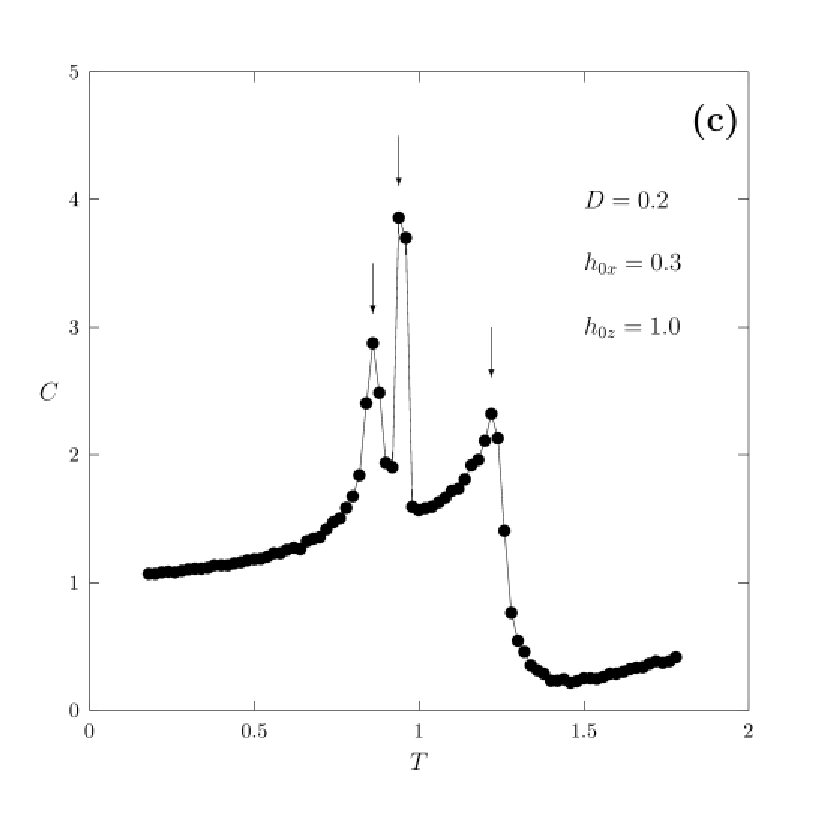}}\\

\caption{ (a): The thermal variations of the dynamic order parameter components illustrated by distinct symbols
$Q_x$ $({\Diamond})$, $Q_y$ $({\circ})$ and $Q_z$ $({\bullet})$.
The amount of the error during the measurement  of $Q_x$, $Q_y$ and $Q_z$ near the critical point is approximately equal to 0.02 and 
that for the lower temperature (i.e., below $T = 0.50J/{k_{B}}$) is approximately equal to 0.003.;(b): The thermal dependence of the nonequilibrium energy.
The critical points are pinned down by vertical arrows(c): The thermal dependence of dynamic specific heat ($C = {{dE} \over {dT}}$). The vertical arrows indicate the position of critical points. These diagram are obtained for constant anisotropy $D = 0.20$ with elliptical polarisation having field amplitude $h_{0x} = 0.30$
and $h_{0z} = 1.00$. Collected from \cite{Acharyya04}}
\label{hpmf}
\end{center}
\end{figure}

The dependency of the different components of the nonequilibrium ordering parameters ($Q_{\alpha}$), upon the temperature,  were investigated and illustrated in Figure \ref{hpmf}(a). When the system undergoes gradually through a colling process from a high temperature  that corresponds to disordered ($\vec Q =0$ ) state, first an intriguing transition occurs which corresponds to an {\it off axial} transition (\cite{Acharyya04}).  The dynamical disordered phase denoted as ($\vec Q=0$,$Q_x=0,Q_y=0,Q_z=0$) shifts to a dynamical {\it Y ordered} phase (P1) specified by P1: ($Q_x=0, Q_y\ne 0, Q_z=0 $) at critical temperature $T_{c1}$. The occurrence of an off-axial transition disrupts the dynamical symmetry of the system, specifically along the y direction. This disruption is a result of applying a magnetic field perpendicular to the preferred axis of order, which, in this case, confined in the plane formed by X and Z axes. When the sample undergoes through the cooling, it cherishes this specific non-equilibrium ordered phase (P1) over an appreciable temperature range.  In the subsequent dynamic phase, denoted as P2, an {\it axial} transition occurs where the ordering characteristics change. Specifically,  the components dynamic order-parameter  are $Q_x \ne 0$, $Q_y=0$ and $Q_z \ne 0$ and this transition takes place at a critical temperature $T_{c2}$. In phase P2, the system exhibits ordering in the plane defined by the applied magnetic field (X-Z plane), while the ordering across the Y direction diminishes. This transition signifies a shift in the dominant ordering behavior within the system.As the temperature continues to decrease, $Q_x$ and $Q_z$ progressively increase.  \\

Upon reaching a specific temperature threshold ($T_{c3}$), another distinct transition occurs, manifesting a striking change in the system's ordering characteristics. In this newly emerged phase P3, the system undergoes a decline in X-ordering while the Z-ordering exhibits a rapid increase. Phase P3 is characterized by P3 : ($Q_x\ne 0, Q_y = 0, Q_z\ne 0 $). Despite their apparent similarities in dynamic order parameter values, P2 and P3 exhibit a crucial distinction. In phase P2, the quantities $Q_x$ and $Q_z$ demonstrate an upward trend as temperature decreases, whereas in phase P3, $Q_x$ exhibits a decline as temperature decreases (see Figure \ref{hpmf} (a)). These distinctions play a significant role in differentiating between the two phases. In phase P3, the system exhibits axial dynamical ordering along the Z-axis. As temperature decreases, the dynamical Z-ordering continues to increase. At very low temperatures, the system enters a phase with exclusive dynamic ordering along the Z-direction (where $Q_x=Q_y=0$ and $ Q_z=1.0 $) , aligning with the direction of anisotropy.

 The thermal dependence of the system's energy (E) is illustrated in Figure \ref{hpmf}(b) to ascertain dynamic transitions and determine their corresponding transition temperatures. The plot clearly shows the three distinct dynamic transitions in this framework, characterized by arc points in the energy-temperature curve. The thermal gradient of energy yields the  dynamical specific heat (C) which is graphically illustrated in Figure \ref{hpmf}(c). The peak observed in the nonequilibrium specific heat plot against temperature (C-T) curve evidently indicate the presence of the three dynamic transition. We can precisely measure the critical temperature by identifying the locus of the peaks: $T_{c1}=1.22$ for first transition (P1), $T_{c2}=0.94$ for the second transition (P2) and $T_{c3}=0.86$ for the last one (P3).

\newpage

\newpage

 The scenario of multiple transitions evanesces for higher amplitude of field strength $h_{0x}$ ( while keeping the other parameters fixed). The C-T curve exihibits two peaks for $h_{0x}=0.9$, indicating the presence of two distinct transitions. The second phase P2 is no longer observed in this case. The system dynamically orders only along the direction of anisotropy (Z direction) and evidences a single transition for for $h_{0x}=h_{0z}=0.2$ [circularly polarized; remaining all other parameter constant].

 The system endures multiple dynamic phase transition for specific ranges of field amplitudes and for lower values of anisotropy.  The Landau-Lifshitz-Gilbert equation accompanied by Langevein dynamics can be used to substantiate this phenomena. The synchronized rotation of spins may originate the multiple (longitudinal and transverse) dynamic transition  in classical Heisenberg system having anisotropy. On the other hand in  discrete model like Ising Model, the phenomena  is commonly elucidated through the process of spin reversal and nucleation. However more detailed analysis are required to confirm the accountable reason for the appearance of multiple phases in non-equilibrium system.
 
\begin{figure}[h]
\begin{center}
\resizebox{8cm}{!}{\includegraphics[angle=0]{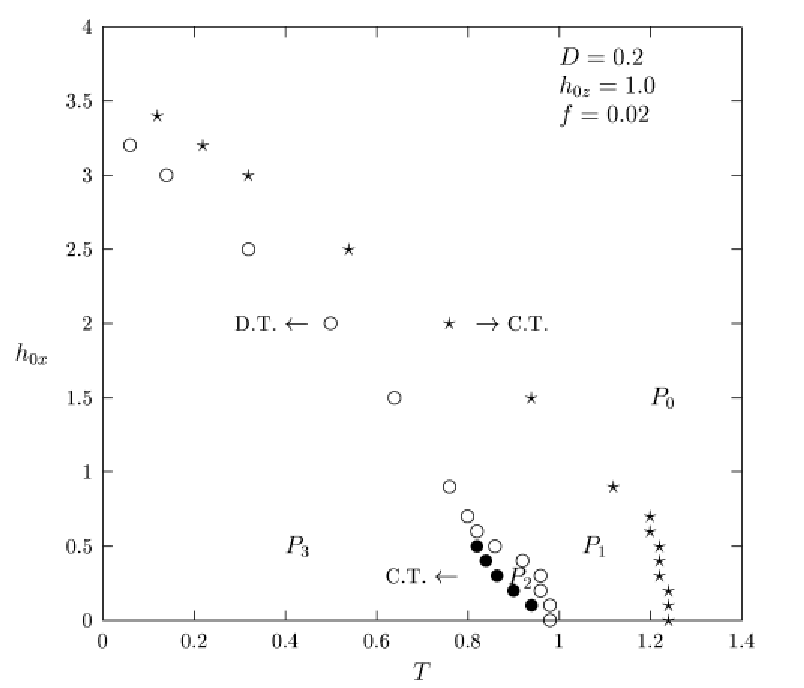}} 
\caption{The phase diagram illustrates nonequilibrium multiphasal transitions in the $T-h_{0x}$ plane for parameters D=0.2, f=0.02, and $h_{0z}$=1.0. The diagram distinguishes between continuous transitions (C.T.) and discontinuous transitions (D.T.) across the phase separating lines, indicating the nature of the transitions. Collected from \cite{Acharyya05}}
\label{phase}
\end{center}
\end{figure}

The phase diagram, depicted in Figure. \ref{phase}, exhibits the regions of different dynamic phases as a function of the applied field amplitude ($h_{0x}$) and temperature ($T$). Multiple dynamic transitions occur, giving rise to a complex phase landscape.
In the  Figure. \ref{phase}, the superficial boundary separates the dynamically ordered phase ($P_1$:Y ordered) specified by $Q_x = 0$, $Q_y \neq 0$, and $Q_z = 0$ from the disordered phase ($P_0$) characterized by $Q_x = 0$, $Q_y = 0$, and $Q_z = 0$. The transition temperature decreases with increasing field amplitude ($h_{0x}$).
For small values of $h_{0x}$, a distinct region confined by the symbols circle and bullet corresponds to the ordered phase denoted as $P_2$. In this phase, $Q_x \neq 0$, $Q_y = 0$, and $Q_z \neq 0$. Although this region is small in area, it is clearly observed and carefully analyzed. As the field amplitude $h_0x$ grows, the transition temperatures for both extends of this phase drop.\\

The region inside the phase plane, bounded by both the circle (representing large $h_{0x}$ values) and the bullet (representing small $h_{0x}$ values), corresponds to the low-temperature phase labeled as $P_3$. In this phase, $Q_x$ and $Q_y$ are zero, while $Q_z$ is non-zero. Analogous to the other phases, the transition temperature exhibits a reduction with an increase in the field amplitude $h_{0x}$.
The obtained phase diagram provides valuable insights into the complex behavior of the system under varying temperatures and applied field amplitudes. The different dynamic phases and their boundaries offer a comprehensive understanding of the interplay between temperature, field amplitude, and the resulting magnetization patterns.

The study investigates the nature of multiple dynamic transitions in a system by analyzing the distributions of dynamic order parameter components near the transition points. The transitions are examined in a phase portrait confined in the $h_{0x}-T$ plane. The transition from the high temperature structureless phase ($P_0$) to the $Y$-ordered phase ($P_1$) is found to be continuous. The distribution of $Q_y$ shifts gradually towards zero as the temperature increases, indicating a continuous transition at around $T=1.20$. In contrast, the transition from $P_1$ to $P_3$ is discontinuous. As the temperature decreases, the value of $Q_y$ drops abruptly from a nonzero value to zero. The distribution of $Q_y$ exhibits two peaks near the transition temperature ($T \approx 0.79$), indicating a discontinuous transition. The transition from $P_2$ to $P_3$ is found to be continuous, with the value of $Q_z$ changing smoothly. The Binder cumulant ratio $U_y$ confirms these findings, showing a monotonic change for the continuous transition and a sharp minimum for the discontinuous transition. The nature of the transitions is summarized in a phase diagram, providing a comprehensive representation of the system's dynamic behavior.

The study also highlights the importance of choosing appropriate field amplitudes for accurate analysis. The chosen value of $h_{0x} = 0.7$ allows for well-separated phase boundaries, facilitating the examination of temperature variations. Furthermore, the analysis of the distribution of $Q_z$ near the transition from $P_2$ to $P_3$ proves challenging due to the small change in $Q_z$. However, the thermal dependence of the Binder cumulant ratio $U_z$ shows a smooth variation, indicating a continuous transition.

Overall, the results demonstrate the existence of both continuous and discontinuous transitions in the system. The study provides valuable insights into the dynamic behavior and nature of these transitions, contributing to a better understanding of complex systems. Further investigations, including larger system sizes and precise estimation of transition temperatures, are suggested to study scaling behavior and critical exponents associated with these transitions.

The findings of a finite size study conducted and briefly documented \cite{Acharyya05} suggest that the observed multiple non-equilibrium transitions presented earlier are unlikely to be a result of finite size effects.

The thermal dependence of  the variances of the non-equilibrium order parameter components ($Q_x$, $Q_y$, $Q_z$) for various system sizes are studied. The results show that the variances exhibit sharp peaks at the transition temperatures ($T_{c1}$ and $T_{c2}$) corresponding to the dynamic phase transitions. The height of the peaks increases with larger system sizes, indicating the presence of a diverging length scale at the transition points. The study also suggests that the transition temperature ($T_{c3}$) for the transition from $P_2$ to $P_3$ phase cannot be resolved from the present analysis. The error bars in the components of the ordering parameters show maximum growth in the vicinity of $T_{c1}$ and $T_{c2}$, while the behavior near $T_{c3}$ is less clear. These findings provide valuable insights into the dynamic behavior of the system and highlight the importance of system size in understanding the transition phenomena.


A brief investigation was conducted on the frequency dependence of the multiphase nonequilibrium transitions.  The thermal dependece of  the components of the non-equilibrium order parameter are studied for a small frequency ($f = 0.01(MCSS)^{-1}$) of the polarized magnetic field, the transitions were studied under the conditions of $D=0.20J$, $h_{0x}=0.30J$, and $h_{0z}=1.0J$. Similar temperature trends were observed compared to the $f=0.02(MCSS)^{-1}$ case ($D=0.2J$, $h_{0x}=0.3J$, and $h_{0z}=1.0J$), albeit with a noticeable shift of the critical points towards lower temperatures. By examining the thermal variations of the ${{dQ_x} \over {dT}}$, ${{dQ_y} \over {dT}}$, and ${{dQ_z} \over {dT}}$, the approximate values of $T_{c1} \simeq 1.06$, $T_{c2} \simeq 0.76$, and $T_{c3} \simeq 0.62$ were obtained. A similar analysis was performed for comparetively a large values of $h_{0x} (=1.5)$ assuming $f=0.01$, $D=0.2$, and $h_{0z}=1.0$. In this scinario, only two transitions were obtained, and the estimated critical temperatures were $T_{c1} \simeq 0.60$ and $T_{c3} \simeq 0.30$. These investigations provide evidence that as the frequency decreases, approaching the equilibrium limit, the critical points in the phase diagram shift towards lower temperatures while maintaining the general structure of the diagram unchanged (for $f=0.02$). This observation aligns with the expectation that at lower frequencies, the magnetization components have more time to align with the field, resulting in the disappearance of the dynamic transition. Hence, this transition is referred to as a "truly dynamic" transition.



What will be the responses of a needle like nanomagnet to the electromagnetic (EM) radiation passing through it? To study
such kind of nonequilibrium responses, 
the needle like anisotropic Heisenberg ferromagnet irradiated by the
propagating electromagnetic field wave has been studied by standard MC simulation using
Metropolis rate of single spin-flip. The electromagnetic wave has been considered to propagate through the
z-axis (easy axis here). The needle was found to show a 
nonequilibrium kind of phase transition at a finite critical temperature where time averaged magnetisation over the full cycle of the propagating radiation assumes nonzero value in the low temperature ordered phase.

The comprehensive phase diagram has been shown in the Euclidian plane formed by the critical temperature and the maximum magnitude of the spatio-temporal variation of the 
propagating magnetic field wave. The following figure (Fig-\ref{rodlike}) shows the comprehensive phase diagram.

\begin{figure}[h]
\begin{center}
\resizebox{8cm}{!}{\includegraphics[angle=0]{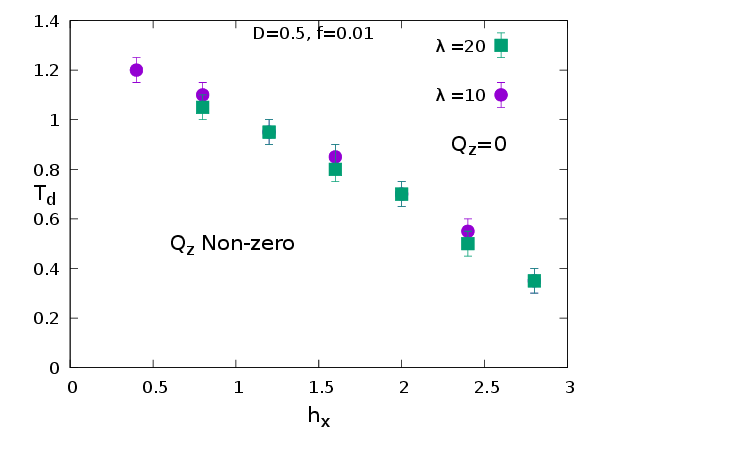}}
\resizebox{8cm}{!}{\includegraphics[angle=0]{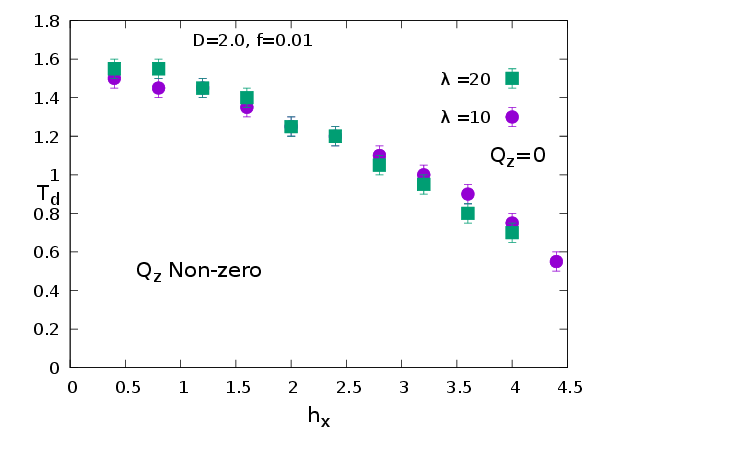}}
\caption{The phase separating boundaries for the phase transition of nonequilibrium type in rodlike Heisenberg ferromagnet for different anisotropies
($d=0.5$(left) and $D=2.0$ (right)) Dfferent symbols represent  wavelengths ($\lambda$) of different values of travelling magnetic field wave. Collected from \cite{Acharyya22}}
\label{rodlike}
\end{center}
\end{figure}

\newpage

\section{Technological significance:}

Magnetism has taken an important place in modern technology. Magnetic force is one of the most fundamental forces in the universe and has been used by mankind for thousands of years. The magnetic memory device is the most common example of the use of magnetic materials. In present days, the great
challenge is to design such a memory or recording device which maximizes the storage capacity and recording speed. To develope such materials one
should have adequate knowledge about its magnetic properties along with the thermodynamical behaviours. Theoreticians play the significant role to
understand the various thermodynamic phases through model calculations. 
The computer simulational studies helps to predict the behaviours in the
limiting points. Moreover, the microscopic configuration or the morphological structures of magnetic materials, can be visualised through
the computer simulation of such classical vector (spin) models. 

The magnetic materials having very large values of magnetic moments can be a special respesentation of classical spin vectors. These kinds of materials can be used in {\it magnetic coating} devices. These devices are operationally active in specially desinged optically sensitive domain. The low temperature
behaviours of such systems can be a realisation of classical Heisenberg or planar ferromagnet irradiated by electromagnetic
field wave. The domain of maximal activity may be searched in the symmetry broken dynamically ordered phase as described in
Section-3. Moreover, the widely used SMOKE (Surface magneto optic Kerr effect) can be employed to study the effective change
in the magnetisation in the dynamically symmetric phase, which may help to identify the maximally active domain as stated
above.

\section {Summary:}

It should be mentioned here that this article is mainly reviewed the equilibrium and nonequilibrium 
phase transitions studied in classical XY and Heisenberg models. The quantum behaviour of these models show many interesting behaviours which is beyond the scope of present article. The interested reader may consult the book \cite{Sachdev2000} Quantum Phase Transitions by Subir Sachdev. The quantum phase transitions \cite{Dutta15} are mainly studied by tuning the transverse field. The thermodynamic behaviours of such magnetic models are not the main goal of such studies.

The thermodynamic behaviours of classical XY and Heisenberg model are briefly reviewed here. In this view, the interested reader may get all necessary up-to-date informations in this review article. Some recent studies are also included here. The main systems of review are XY model and Heisenberg model. The whole article is mainly divided in two parts: the equilibrium behaviours and the nonequilibrium behaviours. The nonequilibrium behaviours are modelled by the time dependence of the generic Hamiltonian through a time dependent externally applied magnetic field. On the other hand the equilibrium behaviours are thoroughly reviewed on the basis of the roles of anisotropy, dilution as the effects of diisorder.

In the study of equilibrium type of phase transitions, the dependences of the critical temperatures on the amount of disorder
(anisotropy, dilution etc.) has been discussed with recent developements. The quantum and classical calculations are
discussed in this context.
 In some cases, the distributed anisotropy was found to play a partinent role in determining the critical temperature. 
 The most recent results are discussed in details. The thermodynamics behaviours of metamagnets (layered antiferromagnets), modelled by classical XY and Heisenberg systems, are also discussed (with phase diagrams) in this article. In some cases,
 for completeness, the experimental results and the theoretical results are compared with reasonable agreement.
 
On the other hand,  nonequilibrium phase transitions in XY and Heisenberg ferromagnets, investigated in last two decades,
have been discussed in details. The nonequilibrium phase transitions, in such systems, driven by oscillating, propagating,
polarised magnetic fields are discussed. The nonequilibrium phase transition in XY nanorod is also analysed.

A section (section-3) has been devoted here to the technological significance of such theoretical studies. Basically,
the metamagnetic behaviours of such systems (In ${\rm FeBr_2}$) have been succesfully modelled by Heisenberg metamagnets.
The rich phase diagram of such systems may be employed to design magnetic
materials for interesting magnetocaloric effects..

Lastly, we believe that this review is a modern documentation of the thermodynamic behaviours and ordering of various magnetic systems which can
be modelled by classical XY and Heisenberg ferromagnets. This will be useful for the modern research in the field of
thermodynamics of magnetism and its statistical aspects in the context of quenched disorder.

\vskip 0.5 cm

\section {Acknowledgements:} The MANF, UGC, Govt. of India is gratefully acknowleded by Olivia Mallick for financial support. Muktish Acharyya
thankfully acknowledges FRPDF, Presidency University. Collaborations with Ulrich Nowak, Klaus Usadel and Erol Vatansever are gratefully acknowledged. We acknowledge Ishita Tikader for a careful reading of the manuscript. We sincerely thank Dr. Anuj Bhowmick for his kind help in preparing the manuscript.


\vskip 1cm

\bibliographystyle{apalike}
\bibliography{reference}


\end{document}